\documentclass[aip,amsmath,amssymb,reprint,twocolumn]{revtex4-2}
\usepackage{graphicx}
\usepackage{subfigure}
\usepackage{adjustbox}
\usepackage{bm}
\usepackage{bbm}
\usepackage{color} 
\usepackage{soul}

\usepackage[dvipsnames]{xcolor} 
\usepackage{braket}
\usepackage{standalone}
\usepackage{multirow}
\usepackage{tikz}
\usepackage{mathrsfs}
\usepackage{dsfont}
\usepackage[colorlinks,bookmarks=true,citecolor=blue,linkcolor=blue,urlcolor=blue]{hyperref}
\usepackage{cleveref}
\usepackage{comment}
\usepackage{mathtools}

 \Crefname{equation}{Eq.}{Eqs.}
\Crefname{figure}{Fig.}{Figs.}

\begin{document}

\title{Non-Hermitian extended midgap states and bound states in the continuum}

\author{Maria Zelenayova}\email[Author to whom correspondence should be addressed: ]{maria.zelenayova@fysik.su.se}
\affiliation{Department of Physics, Stockholm University, AlbaNova University Center, 10691 Stockholm, Sweden}
\author{Emil J. Bergholtz}\email[Author to whom correspondence should be addressed: ]{emil.bergholtz@fysik.su.se}
\affiliation{Department of Physics, Stockholm University, AlbaNova University Center, 10691 Stockholm, Sweden}

\date{\today}

\begin{abstract}
We investigate anomalous localization phenomena in non-Hermitian systems by solving a class of generalized Su-Schrieffer-Heeger/Rice-Mele models and by relating their provenance to fundamental notions of topology, symmetry-breaking and biorthogonality. We find two types of bound states in the continuum, both stable even in the absence of chiral symmetry: the first being skin bulk states, which are protected by the spectral winding number. The second type is constituted by boundary modes associated with a quantized biorthogonal polarization. Furthermore, we find an extended state stemming from the boundary state that delocalizes while remaining in the gap at bulk critical points. This state may also delocalize within a continuum of localized (skin) states. These results clarify fundamental aspects of topology and symmetry in the light of different approaches to the anomalous non-Hermitan bulk-boundary correspondence, and are of direct experimental relevance for mechanical, electrical and photonic systems.

\end{abstract}
\maketitle

It is standard textbook knowledge that states within the energy continuum are generally extended while states in the gap are spatially localized. 
However, almost a century ago, von Neumann and Wigner challenged this paradigm \cite{Wigner1929}, by proposing a bound state with an energy embedded within the extended states, hence providing an example of an anomalous localization phenomenon. This work was extended in the 1970s by Stillinger and Herrick \cite{Stillinger1975}, and briefly addressed in the following years \cite{Friedrich1985,Friedrich1985B,Sadreev2006}. In the past decade, however, there has been renewed interest in bound states in the continuum (BICs) across different physics communities  \cite{Hsu2016, Koshelev2021, Sadreev2021, BoHsu2014, Xu2023}, featuring several potential applications in electronics \cite{Koshelev2018, Kupriianov2019, Calajo2016A}, mechanical metamaterials \cite{Wang2022B}, plasmonics \cite{Wang2022C,Wang2023B, Xiao2020B, Liang2020,Aigner2022} and notably in photonics \cite{Kodigala2017, Longhi21, Hu2022, Qin2022, Kang2023}.

In Hermitian systems, BICs were realized in optical set-ups using trapped light \cite{Hsu2013theory, Hsu2013experiemnt}. Topological properties of BICs in the form of vortex centres in the polarization have been proposed \cite{BoHsu2014}, experimentally realised \cite{Doeleman2018}, and used for unidirectional guided resonances \cite{Yin2020}. BICs can also be induced by moving flat bands \cite{Huang2022} or topological edge states \cite{Zhang2021,Tao2016,Calajo2016B}.
In non-Hermitian systems, impurity-induced BICs were studied \cite{Poli2015,Fang2023} or recently, a generalized framework to study BICs in nanophotonics lattices with engineered dissipation was introduced in Ref.~\onlinecite{Gong2022}. 

In the context of non-Hermitian topological phenomena \cite{Bergholtz2021}, localization phenomena go beyond the single bound state in the continuum. In particular, the non-Hermitian skin effect (NHSE) \cite{Yao2018, Lee2016, Okuma2020, Kunst2018, Martinez2018, Xiao2020, okuma2023, Lin2023} and biorthogonal bulk boundary correspondence when\cite{Kunst2018} viewed as localization phenomena provide a natural framework to study the interplay of localized and extended states. Such states give rise to several anomalous localization phenomena with no known counterpart in the Hermitian systems. In addition to a single bound state, a continuum of bound states was also proposed \cite{Wang2023}. In contrast to localized states, an extended mid-gap state (EGS) was realized \cite{Ghatak2020, Wang2022} and analysed \cite{Schomerus2020} in metamaterials, and proposed using acoustoelectric effect \cite{Gao2020}. An antipode of the bound states in the continuum, an extended state lying inside the continuum of bound states (ELC), was proposed in Ref. \onlinecite{Wang2022B}. Anomalous localization phenomena is also known to lead to a remarkable spectral sensitivity \cite{Schomerus2020,Budich2020,Edvardsson2022}, which may be harnessed in sensing devices \cite{Budich2020,konye2023nonhermitian,electricsensor,parto2023enhanced}. 

The aforementioned anomalous (de)localization phenomena, have been dispersed throughout various models and platforms. In this Letter, we (i) unify all these (de)localization phenomena by demonstrating their emergence in a single general one-dimensional non-Hermitian model realizable in several platforms; (ii) show that biorthogonal polarization ${\mathcal P}$ predicts the gap closing and emergence of bound states in the continuum and remains relevant predictive quantity in the absence of chiral symmetry; (iii) put these results in the context of non-Hermitian topology and the anomalous bulk-boundary correspondence of non-Hermitian systems.

In what follows, we show BIC emerging when the  ${\cal P}$ changes its values and the robustness of  ${\cal P}$ in the broken chiral symmetry phase. In addition to BIC, viewing the skin states as the localized continuum give rise to the bound states in the continuum of bound states (BICB). Contrary to localized states, we demonstrate that EGS and ELC require the boundary state to become extended at the critical point.

We consider the open non-Hermitian Su–Schrieffer–Heeger chain (SSH) with a nonreciprocal $\gamma$ term in the intra-cell hopping. The SSH chain is placed in the staggering potential $\Delta$ breaking the chiral symmetry and turning it into a non-Hermitian Rice-Mele model,
\begin{eqnarray}
    H & = & \sum_{n=1}^{N_{\rm cell}-1} \big [ (t_1 - \gamma) a^{\dagger}_{n,B} a_{n,A}+(t_1 + \gamma) a^{\dagger}_{n,A} a_{n,B}  \nonumber \\
    & + & t_2 (a^{\dagger}_{n,B}a_{n+1,A}+a^{\dagger}_{n+1,A}a_{n,B}) +\Delta (n_{n,A} - n_{n,B}) \big ] \nonumber \\ 
    & + & t_2 (a^{\dagger}_{{N_{\rm cell}-1},B} a_{{N_{\rm cell}},A}+a^{\dagger}_{N_{\rm cell},A} a_{{N_{\rm cell}-1},B}) \nonumber \\
    &+&\Delta n_{N_{\rm cell},A}
    \label{eq:main_model},
\end{eqnarray}
where $N_{\rm cell}$ is the number of cells, $n_{n,\alpha}=a_{n,\alpha}^{\dagger} a_{n,\alpha}$ is the number of particles, and $a^{\dagger}_{n,\alpha} (a)$ creates (annihilates) a bosonic particle in a unit cell $n$ in a sublattice $\alpha$. The open boundaries are applied such that the last unit cell is broken, and the model has $2N_{\rm cell}-1$ sites in total. 

For our purposes, it is crucial to consider a chain with an odd number of sites. This creates an "anomaly" where the boundary mode accounts for the difference between the $A$ and $B$ sublattices by being confined entirely to the  $A$ sublattice. This inhibits hybridization between boundary and bulk modes and thus facilitates their independent tuning, which is reflected in the structure of the energy spectrum. Generalizing Refs. \onlinecite{Edvardson2020, KunstMiertBerg2019, 
 KunstMiertBerg2018, Edvardsson2022thesis,Kunst2018,KunstDwivedi2019,Fan2022Liov} we find
 \begin{equation}
   \{ E_{-}(k), \Delta, E_{+}(k) \},
 \end{equation} where the bulk energy spectrum is derived from the Hamiltonian with periodic boundaries by shifting the momentum $E (k) = E^{\rm PBC} (k - i  {\rm \log} ( r ) )$, where $r~=~\sqrt{t_1-\gamma} / \sqrt{t_1+\gamma}$  \cite{Fan2022Liov}. Hence, we find that
\begin{equation}
    E_{\pm}(k) = \pm \sqrt{ t_1^2+t_2^2 + \Delta^2 - \gamma^2 + 2 t_2\sqrt{t_1-\gamma} \sqrt{t_1+\gamma} \cos{k} }
    \label{eq:main_energy}
\end{equation}
where $k=\pi m/N_{\rm cell} $, with $ m =1,2,...(N_{\rm cell}-1)$.

\begin{figure*}[!]
\includegraphics[width=0.99\textwidth]{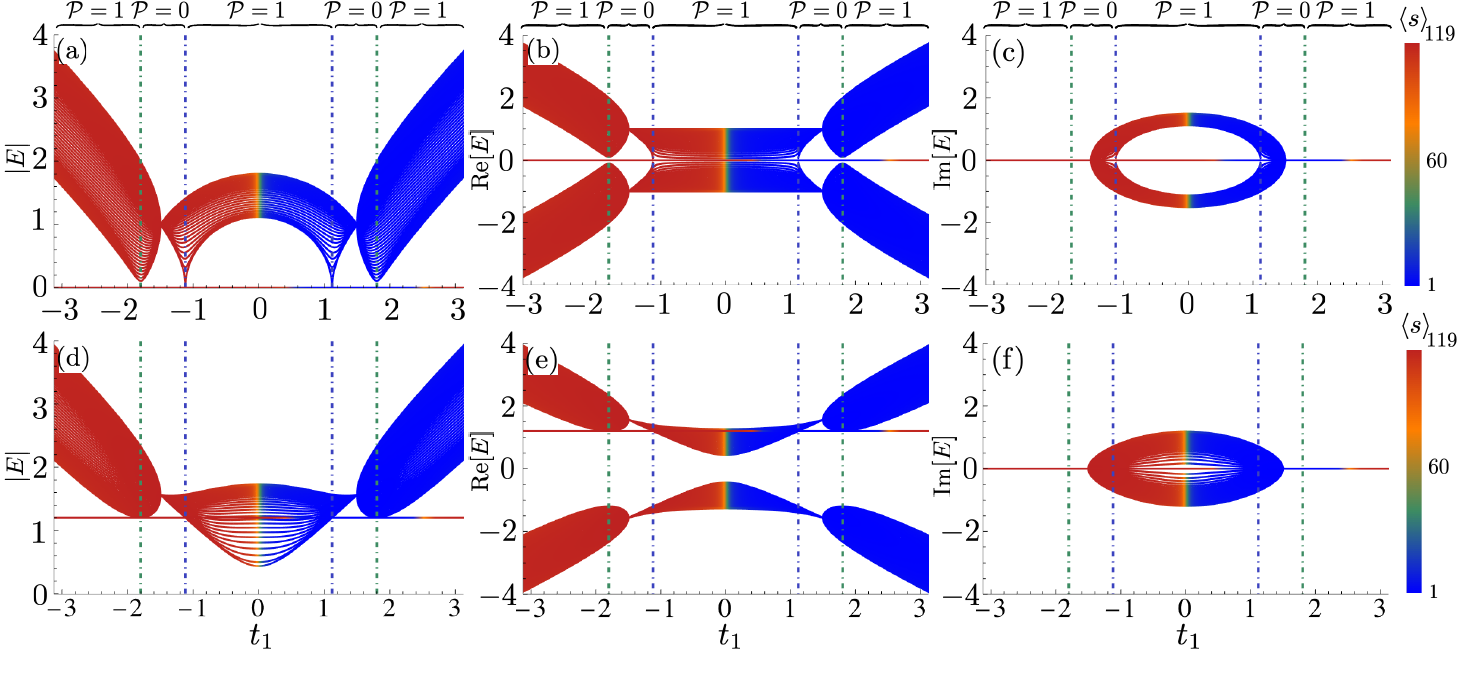}
\caption{The energy spectrum of the non-Hermitian SSH chain, Eq.~(\ref{eq:main_model}) with $t_2=1$, $\gamma=3/2$, $N_{\rm cell}=60$ for $\Delta=0$ and $\Delta=1.2$ (Rice-Mele model) in the top, (a)-(c), and bottom (d)-(f) panels, respectively. The coloring of the energy points is determined by the average position  $\left < s \right>$, of the corresponding mode as mentioned in the main text. Focusing on the energy of the edge mode, Eq.~(\ref{eq:exact_zero_mode}), lying at $E=\Delta$, the figure illustrates i)the change of the localization boundary from the right (red) to the left (blue) at $t_1$ given by Eq.~(\ref{eq:condition_extended}); ii) at the transition point, Eq.~(\ref{eq:condition_extended}), the edge mode becomes extended, see (e) and (f) in Fig.~\ref{fig:states} for comparison; iii) the gap closes where ${\cal P}$ changes, Eq.~(\ref{eq:P_change_points}), marked by the green and blue dash-dotted lines at $r_l^*r_r=1$ and $r_l^*r_r=-1$, respectively. For the latter, polarization also marks the emergence of BIC (BICB).}
\label{fig:energy}
\end{figure*}

Fig.~\ref{fig:energy} shows the energy spectrum as a function of intra-cell hopping $t_1$ for all $k$ values, where the color of the energy point represents the average position $\left< s \right > $ of the particle described by the right eigenstate $\Psi_{R,{\rm Bulk}}(k,n)$,  hence $ \left < s \right > = \sum_{n,\alpha=A,B}^{N_{\rm cell}} s_{n,\alpha} |\Psi_{R,{\rm Bulk}}(k,n)|^2$,  where $s_{n,\alpha}$ yields the site label of the sublattice $\alpha $ in the unit cell $n$, $s_{n,\alpha}=2n-\delta_{A, \alpha}$. 

While the periodic Bloch energies of our model lack spectral inversion symmetry, the open boundary energies (Eq.~(\ref{eq:main_energy})) have the symmetry $E(k)=E(-k)$ which makes it possible to derive all eigenstates generalizing Refs.~\onlinecite{Edvardson2020, KunstMiertBerg2019}. The right bulk modes are
$\Psi_{R,\mathrm{Bulk},\pm(k)}$ and read
\begin{equation}
\Psi_{R,\mathrm{Bulk},\pm}(k) = \mathcal{N} \begin{pmatrix}
	\Psi_{R,\mathrm{Bulk},\pm,A}(k,1) \\
	\Psi_{R,\mathrm{Bulk},\pm,B}(k,1)\\
	\Psi_{R,\mathrm{Bulk},\pm,A}(k,2)\\
	\vdots\\
	\Psi_{R,\mathrm{Bulk},\pm,A}(k,n)\\
	\Psi_{R,\mathrm{Bulk},\pm,B}(k,n)\\
        \vdots\\
	\Psi_{R,\mathrm{Bulk},\pm,B}(k,N_{\rm cell}-1)\\
	\Psi_{R,\mathrm{Bulk},\pm,A}(k,N_{\rm cell})\\
	\end{pmatrix},
\label{eq:bulk_states_full}
\end{equation}
where $n$ labels the unit cell, and the amplitudes read
\begin{equation}
\begin{split}
	& \Psi_{R,\mathrm{Bulk},\pm, A}(k,n) = r^n \left [ (t_1 \!+\! \gamma) \sin{(k n)}\! +\! \frac{t_2}{r} \sin{(k [n-\! 1])} \right],\\
 & \Psi_{R,\mathrm{Bulk},\pm, B}(k,n) = r^n \left [ E_{\pm}(k)-\Delta \right ] \sin{(k n)},
 \end{split}   
\label{eq:bulk_states_sublattices}
\end{equation}
where $r=\sqrt{\frac{t_1-\gamma}{t_1+\gamma}}$.
From these expressions, we see that states are exponentially localized to one of the boundaries, the phenomenon known as the \textit{non-Hermitian skin effect} (NHSE) \cite{Kunst2018, Martinez2018, Brandenbourger2019}, where the value of the spectral winding number \cite{okuma2023} predicts the boundary to which states pile up. In Fig.~\ref{fig:energy}, the average position  $\left < s \right>$ shows the localization to the right (red) boundary for $t_1<0$ and to the left (blue) boundary for $t_1>0$. Comparing the top and bottom panels with $(\Delta = 0)$ and without $(\Delta \neq 0)$ chiral symmetry, respectively, we see that the NHSE is carried over to the broken chiral symmetry phase.

The remaining eigenstate, the edge mode, with energy $\Delta$ is an exponentially localized boundary mode
\begin{equation}
	\ket{\psi_{R/L}} = \mathcal{N}_{R/L}\sum_{n= 1}^{N_{\rm cell}} r_{R/L}^n a_{n,A}^{\dagger}\ket{0},
 \label{eq:exact_zero_mode}
\end{equation}
where $\mathcal{N}_{R/L}$ is the normalization, and  $r_{R}=-(t_1-\gamma)/t_2$ and $r_{L}^{*}=-(t_1+\gamma)/t_2$ dictate the behaviour of the right ($R$) and left ($L$) boundary modes \cite{Edvardson2020,Kunst2018}. (Note that right/left here refers to the eigenvectors fulfilling $H\Psi_R=E\Psi_R$ and  $\Psi_LH=E\Psi_L$, respectively, rather than where the state may be localized.)

While the emergence of NHSE is related to a spectral winding number \cite{okuma2023}, the behaviour of boundary $\Delta$-mode can be associated with the biorthogonal polarization ${\cal P}$ \cite{Edvardson2020, Kunst2018}, see e.g.  Ref.~\onlinecite{Edvardson2020} for a review of this concept,
\begin{equation}
\mathcal{P} = \lim_{N_{\rm cell} \rightarrow\infty}\frac{1}{N_{\rm cell}} \sum_{n}  \frac{\braket{\psi_L| {n} \Pi_{n}|\psi_R}}{{\braket{\psi_L|\psi_R}} },
\label{eq:original_biorth_pol}
\end{equation}
where $\ket{\psi_{R/L}}$ are the boundary modes in  Eq.~\eqref{eq:exact_zero_mode}, $\Pi_n$ is the projection operator on each unit cell $n$. Note that in the Ref.~\onlinecite{Kunst2018} the biorthogonal polarization is defined as $1-{\mathcal P}$.

Remarkably, the change in ${\cal P}$ implies the gap closings \cite{Edvardson2020, Kunst2018}. To see this, we evaluate the limit in Eq.~(\ref{eq:original_biorth_pol}) and find that the biorthogonal polarization ${\mathcal P}$ changes when $|r_L^* r_R|=1$ \cite{Edvardson2020}, i.e., when

\begin{align}
t_1=\pm \sqrt{\gamma^2 + t_2^2}, && t_1=\pm \sqrt{\gamma^2 - t_2^2}.
\label{eq:P_change_points}
\end{align} Evaluating Eq.~(\ref{eq:main_energy}) for $\lim_{N_{\rm cell} \rightarrow\infty}$ at $t_1$ in Eq.~(\ref{eq:P_change_points}) we find  $E(k=N_{\rm cell}/2)=\Delta$, i.e., the gap between the bulk modes and the $\Delta$-mode closes as predicted by the change in the polarization number ${\cal P}$. This is also visible in Fig.~\ref{fig:energy}, where the colored dashed dotted lines highlight the polarization ${\cal P}$ accompanied by the gap closings, and discrepancies are due to the finite size of the system.

$\mathcal P$ also marks the emergence of the \textit{bound state in the continuum}(BIC). Once the gap is closed, the gap can open again, or remain closed. This is visible in the middle panels of Fig.~\ref{fig:energy}, where the boundary mode at $E=\Delta$ touches the bulk (green lines) or enters the continuum (blue lines) at the points marked by the change of ${\cal P}$.

\begin{figure*}[!]
\includegraphics[width=0.99\textwidth]{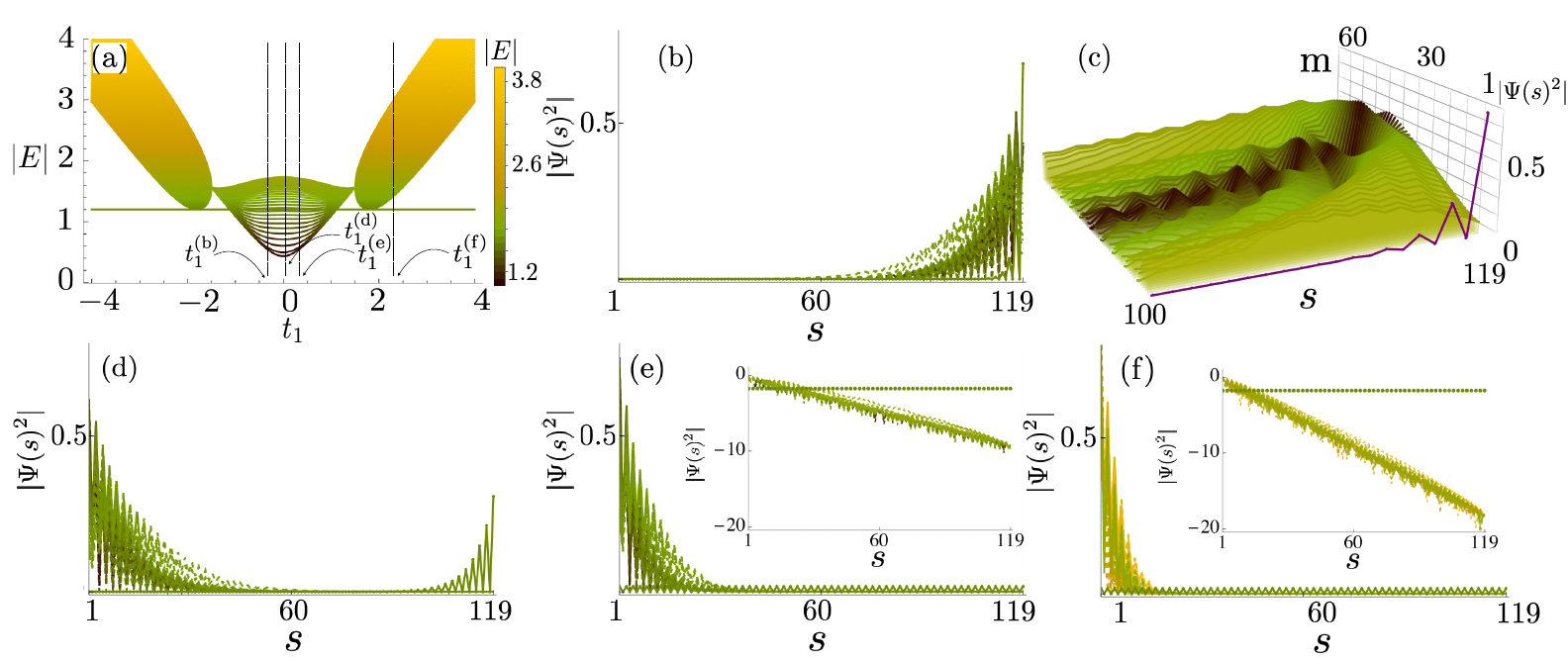}
\caption{
The wave-function squared of the eigenstates in Eq.~(\ref{eq:bulk_states_full}) and Eq.~(\ref{eq:exact_zero_mode}), where $s$ labels the sites, $t_2=1$, $\gamma=3/2$, $\Delta=1.2$, $N_{\rm cell}=60$ and the coloring of the mode corresponds to the absolute value of the energy of that particular mode as shown in (a). Bound states in the continuum at $t_1^{(b)}=-0.3$ and $t_1^{(d)}=0.3$ are shown in (b)  and (d), respectively. For the former, the edge state, highlighted in purple in (c) panel, is localized among the bulk modes, leading to BICB. For the latter, the edge mode localizes on the opposite boundary to the skin states. An extended state at $t_1^{(e)}$ with energy lying inside the localized continuum of bound states (ELC) is shown in (e), while EGS emerging at $t_1^{(f)}$ is visible in (f). The insets in (e) and (f) show the same data with y-axis on a logscale exemplifying the exponential dependence of the skin states and delocalization of the edge state.}
\label{fig:states}
\end{figure*}

The localization boundary of the BIC is determined by the amplitude in Eq.~\eqref{eq:exact_zero_mode} and can be the same or opposite to the localization boundary of the skin states, see Fig.~\ref{fig:states} (b) and (d) respectively.

In Hermitian systems, bound states appear with discrete energies. In comparison, for the non-Hermitian system we consider here, the bulk skin states can form a continuum in the limit that $N_{\rm cell} \to \infty$, leading to \textit{a continuum of bound skin states}. This is visible in the middle panels in Fig.~\ref{fig:energy}, where the bulk energies start to form the continuum. A similar phenomenon was proposed using imaginary momentum and Landau-type vector potential \cite{Wang2023}, while here we demonstrate the phenomenon using a simple non-Hermitian SSH model.

Another anomalous localization composition is from the combination of the boundary mode, Eq.~(\ref{eq:exact_zero_mode}), and the bulk skin states, Eq.~(\ref{eq:bulk_states_full}), where the former localizes in the latter giving rise to \textit{bound state in the continuum made of bound states} (BICB), see (b) in Fig.~\ref{fig:states}. Therefore, in our model, the BICs can be BICB.

We now turn to anomalous delocalization phenomena. From  Eq.~(\ref{eq:exact_zero_mode}), we see that the edge mode can transition from being localized to the left (right) boundary for $|r_R|<1$ ($|r_R|>1$) to the extended state at $|r_R|=1 $ i.e., when 
\begin{align}
    t_{1 \pm}=\gamma \pm t_2.
    \label{eq:condition_extended}
\end{align}
At these points, the otherwise localized boundary state is now extended throughout the lattice while all the bulk skin states are piled on the boundary, see (e) and (f) in Fig~\ref{fig:states}. The energy of these extended states can lie inside or outside of the continuum. 
When it lies in the continuum, an \textit{ extended state in the localized continuum} (ELC) is formed, see panel (e) in  Fig.~\ref{fig:states}, where the localized continuum is formed by the bulk skin states. An analogue of ELC was realized in a mechanical two-dimensional system \cite{Wang2022B} and we show that it can also emerge in a one-dimensional model.
The extended state lying outside of the continuum creates \textit{extended mid-gap state} (EGS), see panel (f) in Fig.~\ref{fig:states}, which is localized at zero for $\Delta=0$, while for $\Delta \neq 0$ it is shifted by $\Delta$. A similar EGS was realised in metamaterials \cite{Ghatak2020, Wang2022}, the transition to EGS was also analysed \cite{Schomerus2020}, and recently a similar phenomenon was addressed in solitons \cite{Blas2023}.

In the biorthogonal basis, the emergence of the extended state is predicted by the change in $\mathcal P$. In particular, for the chain with the odd number of sites (the last unit cell broken), the biorthogonal-bulk boundary\cite{Kunst2018} correspondence implies that at  ${\cal P}$ changing points, Eq.~\eqref{eq:P_change_points}, the biorthogonal boundary mode \cite{Edvardson2020},  $\braket{\psi_L| \Pi_n|\psi_R}$, becomes extended. This delocalization is different from the extended state in the right eigenvector basis, Eq.~(\ref{eq:exact_zero_mode}) at $|r_R|=1$.

In this work we have fully solved a non-Hermitian Rice-Mele model and shown how the localization of bulk and boundary state defy usual expectations. Notably, this highlights bound states in the continuum (BICs) and extended states in the gap (EGSs) as well as in a localized continuum (ELCs). In fact, the localization of non-Hermitian bulk and boundary states can be tuned essentially independently. 

We have also shown how a jump in the polarization marks the emergence (or disappearance) of BICs. This highlights a fundamental difference between the biorthogonal polarization \cite{Kunst2018}, which carries physical information also when the chiral symmetry is broken ($\Delta\neq 0$), and the generalized Brillouin zone winding number \cite{Yao2018} that instead relies on the aforementioned symmetry. This is notable since their predictions coincide in the SSH limit $\Delta=0$. The difference reflects that the (eigenvector) winding number relates to symmetry-protected topology while the polarization reflects a general feature of diverging (biorthogonal) localization lengths at phase transitions.

The simplicity of our model makes it relevant for many experimental set-ups. Notably, the pertinent non-Hermitian SSH model has been realized in mechanical metamaterials\cite{Ghatak2020,Wang2022}, electric circuits \cite{helbig2020,konye2023nonhermitian,electricsensor} and photonics \cite{Xiao2020,parto2023enhanced}. In fact, the EGS was already observed in Refs.~\onlinecite{Ghatak2020,Wang2022}. Using the same methods and observables considered in the aforementioned variety of platforms, one can now explore the rich physics of BIC, EGS, and ELC in these systems by tuning the appropriate parameters following our unified analysis.  

\acknowledgments
 The authors thank Evan P. G. Gale for valuable comments on the manuscript. The authors are supported by the Swedish Research Council (VR, grant 2018-00313), the Wallenberg Academy Fellows program (2018.0460) and the Göran Gustafsson Foundation for Research in Natural Sciences and Medicine.

\section*{AUTHOR DECLARATIONS}
\subsection*{Conflict of interest}
The authors have no conflicts to disclose.
\subsection*{Author Contributions}
\textbf{MZ}: Writing - original draft preparation (lead); Visualization(lead);  Formal analysis (equal); Writing - review \& editing (equal);
\textbf{EJB}: Conceptualization (lead); Funding acquisition (lead); Supervision (lead); Formal analysis (equal); Writing - review \& editing (equal). 
\bibliography{main.bib}

\end{document}